\documentclass{svjour3}

\usepackage{amstext,amssymb}
\usepackage{paralist}
\usepackage{graphicx}
\usepackage{tikz}

\def\cL{\mathcal{L}}

\def\cT{\mathcal{T}}

\usepackage{basics}
\usepackage{local}
\usepackage[bookmarksopen,bookmarksnumbered,colorlinks,citecolor=blue]{hyperref}

\begin{document}

\title{Experiences from Exporting Major Proof Assistant Libraries}

\author{Michael Kohlhase \and Florian Rabe}
\institute{Computer Science, FAU Erlangen-N\"urnberg}
\date{Received: ??? / Accepted: ???}

\maketitle

\begin{abstract}
  The interoperability of proof assistants and the integration of their libraries is a highly valued but elusive goal in the field of theorem proving.
  As a preparatory step, in previous work, we translated the libraries of multiple proof assistants, specifically the ones of Coq, HOL Light, IMPS, Isabelle, Mizar, and PVS  into a universal format: OMDoc/MMT.

  Each translation presented tremendous theoretical, technical, and social challenges, some universal and some system-specific, some solvable and some still open.
  In this paper, we survey these challenges and compare and evaluate the solutions we chose.

  We believe similar library translations will be an essential part of any future system interoperability solution and our experiences will prove valuable to others undertaking such efforts.
\end{abstract}
\setcounter{tocdepth}{3}

\section{Introduction}\label{sec:intro}
 \paragraph{Motivation}
The hugely influential QED manifesto~\cite{qed} of 1994 urged the automated reasoning community to work towards a universal, computer-based database of all mathematical knowledge, strictly formalized in logic and supported by proofs that can be checked mechanically.
The QED database was intended as a communal resource that would guide research and allow the evaluation of automated reasoning tools and systems. This database was never realized, but the interoperability of proof assistants and the integration of their libraries has remained a highly valued but elusive goal.

This is despite the large and growing need for more integrated and easily reusable libraries of formalizations.
For example, the Flyspeck project \cite{flyspeck} build a formal proof of the Kepler conjecture in HOL Light.
But it relies on results achieved using Isabelle's reflection mechanism, which cannot be easily recreated in HOL Light.
And that is only an integration problem between two systems using the same foundation ---
Users of theorem provers often approach (and occasionally exasperate) developers with requests to be able to use, e.g., Coq's tactics together with Isabelle's sledgehammer tool, requests that sound simple to users but are infeasible for developers.

\paragraph{Problem and Related Work}
No strong tool support is available for any such integration.
In fact, merging libraries can hardly even be attempted because we lack satisfactory mechanisms to translate across languages or into a common language.
Even worse, most integration attempts currently falter already when trying to \emph{access} the libraries in the first place.
The libraries consist of text files in languages optimized for fast and convenient writing by human users.
Consequently, highly non-trivial algorithms for parsing, type reconstruction, and theorem proving have been developed to build the corresponding abstract data structures.
This has the effect that for each library, there is essentially only a single system able to read it.
Moreover, these systems' kernels are typically realized as read-evaluate-print interfaces to the foundation, optimized for batch-processing input files, and appear to the outside as monolithic black boxes.
Therefore, any integration requires theorem prover to support \emph{exporting} libraries in formats that can be read by external tools such as other proof assistants, interoperability middleware, or knowledge management services.

Even when exports exist, there are two major problems.
Firstly, most exports contain the elaborated low-level data structures that are suitable for the kernel and not the high-level structure that is seen by the user.
The latter usually corresponds more closely to the informal domain knowledge and is therefore more valuable for reuse.
Secondly, the export code quickly becomes out-of-date as new features are added to the main system.
The only exception are exports that are actively maintained by the developers of the respective theorem prover, but this is rarely the case.

Therefore, there are only a few examples of successful transports of a library between two proof assistants.
Some have been realized as \emph{ad-hoc logic translations}, typically in special situations.
\cite{hol_coq} translates from HOL Light \cite{hollight} to Coq \cite{coq} and \cite{hol_isahol} to Isabelle/HOL.
Both translations benefit from the well-developed HOL Light export and the simplicity of the HOL Light foundation.
\cite{isahol_isazf} translates from Isabelle/HOL \cite{isabellehol} to Isabelle/ZF \cite{isabelle_zf}.
Here import and export aided by the use of the same logical framework.
The Coq library has been imported into Matita, aided by the fact that both use very similar foundations.
The recent Lean system includes APIs that make translations into other formats relatively easy; this was mostly used for independent proof checking and an integration with Mathematica \cite{lean_mathematica}.
In \cite{nuprl_coq}, the Nuprl language was represented in Coq for the purpose of verifying Nuprl proofs.
The OpenTheory format \cite{opentheory} was developed to facilitate sharing between HOL-based systems but has not been used extensively.

Alternatively, one can use \emph{logical framework-based transports}, where the target system is a logical framework.
This approach was used by us in the work we present here.
It is also used by the Dedukti group, e.g., for HOL Light in \cite{holide} and for Coq in \cite{assafphd}.
In principle, the logical framework can serve as a uniform intermediate data structure via which libraries can be moved into other proof assistants.
Such translations were built in \cite{hol_nuprl2} from a representation of HOL in LF to one of Nurpl and similarly in \cite{CHKMR:latinabs:11} for a large set of logics.
But these approaches lived only in the logical framework and lacked a connection to the actual systems and their libraries.
Recently, small arithmetic libraries were transported in this way using Dedukti as an intermediate \cite{dedukti_sharing}.

\paragraph{Contribution}
In~\cite{KR:qed:14}, we proposed a major project of extending representations of proof assistant logics in logical framework and exporting their \emph{entire} libraries into a universal format.
While far from a final QED interoperability solution, it constituted a critical step towards building interoperability and knowledge management applications.

Since then, we and our research group have put this proposal into practice in the context of the OAF project (Open Archive of Formalizations), including some of the biggest existing libraries.
To use resources efficiently, we chose representative theorem provers: one each based on higher-order logic (HOL Light \cite{hollight}), constructive type theory (Coq \cite{coq}), an undecidable type theory (PVS \cite{pvs}), and set theory (Mizar \cite{mizar}), as well as one based on a logical framework (Isabelle \cite{isabelle}) and one based on axiomatic specification (IMPS \cite{imps}).
All six exports were presented individually before: HOL Light in \cite{KR:hollight:14}, Mizar in \cite{IKRU:mizar:11}, PVS in \cite{KMOR:pvs:17}, IMPS in \cite{imps_oaf}, Coq in \cite{MRS:coq:19}, and Isabelle in \cite{KRW:isabelle:19}.
For simplicity, we will refer to these as ``our exports'' in the sequel even though each one was developed with different collaborators.

Here, we report on our experiences with this enterprise.
In the process of these five years of work, we have accumulated a lot of knowledge that is complementary to the individual papers.
Circumstances led us to try several very different approaches in all these exports.
While the previous papers presented the logical details of each export in depth, the present paper abstracts from the technicalities and discusses the general challenges.
It does not subsume the previous papers, and we repeat citations already present in the previous papers only if they are of particular interest here.
Instead, this paper describes in general the problems (solved or remaining), possible solutions, and future priorities concerning these exports.
We focus on the details that are relevant for comparing and evaluating the approaches and try to record and pass on the knowledge/lessons that will help other researchers attempting future proof assistant exports.
Thus, one can see the individual exports as experiments that created data, and the current paper as the one that interprets and draws conclusions from this data.

\paragraph{Overview}
Section~\ref{sec:general} reports on the common aspects of exporting theorem libraires, and Sections~\ref{sec:coq} to~\ref{sec:pvs} discuss system-specific challenges, solutions, and related work. 
Section~\ref{sec:concl} concludes the paper and points out future work. 
The entire text of this paper is new with the exception that the sections on the individual systems also include short high-level summaries of the respective export to make the paper self-contained.



\section{General Considerations}\label{sec:general}
 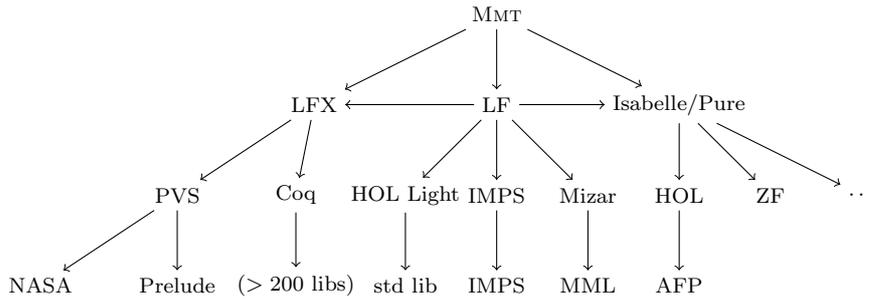
\begin{figure}[ht]\centering
\begin{tikzpicture}[scale=1.2]
\node at (0,0)  (MMT) {\mmt};
\node at (0,-1) (LF)  {LF};
\node at (2,-1) (Pure)  {Isabelle/Pure};
\node at (-2,-1) (LFX)  {LFX};

\draw[->] (MMT) -- (LF);
\draw[->] (MMT) -- (LFX);
\draw[->] (MMT) -- (Pure);
\draw[->] (LF) -- (Pure);
\draw[->] (LF) -- (LFX);

\node at (-3.5,-2) (PVS)  {PVS};
\node at (-2.2,-2) (Coq)  {Coq};
\node at (-1,-2) (HOLL)  {HOL Light};
\node at (0,-2) (IMPS)  {IMPS};
\node at (1,-2) (Miz)  {Mizar};
\node at (2,-2) (IsaHOL)  {HOL};
\node at (3,-2) (IsaZF)  {ZF};
\node at (4,-2) (IsaOther)  {$\cdots$};

\draw[->] (LF) -- (HOLL);
\draw[->] (LF) -- (IMPS);
\draw[->] (LF) -- (Miz);
\draw[->] (LFX) -- (Coq);
\draw[->] (LFX) -- (PVS);
\draw[->] (Pure) -- (IsaZF);
\draw[->] (Pure) -- (IsaHOL);
\draw[->] (Pure) -- (IsaOther);

\node at (-3.5,-3) (PVSPrel)  {Prelude};
\node at (-5,-3) (PVSNasa)  {NASA};
\node at (-2.2,-3) (CoqLib)  {($>200$ libs)};
\node at (-1,-3) (HOLLLib)  {std lib};
\node at (0,-3) (IMPSLib)  {IMPS};
\node at (1,-3) (MizLib)  {MML};
\node at (2,-3) (AFP)  {AFP};

\draw[->] (PVS) -- (PVSPrel);
\draw[->] (PVS) -- (PVSNasa);
\draw[->] (Coq) -- (CoqLib);
\draw[->] (Miz) -- (MizLib);
\draw[->] (HOLL) -- (HOLLLib);
\draw[->] (IMPS) -- (IMPSLib);
\draw[->] (IsaHOL) -- (AFP);
\end{tikzpicture}
\caption{Major Libraries in \mmt as a Universal Framework}\label{fig:share}
\end{figure}

\subsection{General Approach}

We use OMDoc \cite{omdoc} as the standard XML format for holding the library exports.
Its semantics and API are provided by the \mmt system \cite{RK:mmt:10,rabe:mmtabs:13}.
\mmt also provides implementations of logical frameworks such as LF \cite{lf} and extensions (collectively named LFX; see~\cite{DMueller:phd:19}) we have developed for representing theorem prover logics adequately.
With that, the general pattern for exporting the library $\cL$ of a theorem prover $\cT$ is as follows:
\begin{enumerate}
\item Within \mmt, we choose a logical framework in which we can define the logic underlying $\cT$ as a theory $\cT^*$.
  Preferably,   this is LF but in practice often a stronger, sometimes custom-designed framework is needed.
\item We instrument $\cT$ to export a \textbf{system-near rendering} $\cL'$ of $\cL$, usually using a standard format like XML or JSON but with a  system-specific, sometimes ad hoc, schema.
\item We import $\cL'$ into \mmt and
  \begin{compactitem}
    \item standardize or mimic its statement syntax in \mmt,
    \item represent its expression syntax relative to $\cT^*$.
   \end{compactitem}
  \item We serialize the result as OMDoc yielding $\cL^*$.
\end{enumerate}
The resulting collection of libraries is shown in Figure~\ref{fig:share}.

Crucially, this enables a separation of skill sets:
The first step can be best executed by an expert on the logic of $\cT$ (and LF/LFX), and the second step by an expert for system $\cT$, often a core developer.
The latter is required as no system is documented well enough to make an export feasible without a expert on the inside.
The third step can be best executed by an \mmt expert given a direct line of communication to the $\cT$ expert about the meaning and details of $\cL'$.
A good practice is for the two of them to jointly write or document an XML or JSON schema for the system-near export.
A two-week research visit proved to be the minimal investment necessary to synchronize this part and to specify (not implement) the export.

The separation above allows maintaining the generation of $\cL'$ together with the $\cT$ release process and that of $\cL^*$ with the \mmt release process.
This has proved to be the best recipe against the bit-rot of exporters, which is one of the biggest problems in practice.

Even so, the involved tasks are badly incentivized: they are very difficult, very work-intensive, and tend to produce brittle implementations that generate only one paper.
This is an open problem with academic system development in general and prover library exports in particular --- our exports were only possible due to well-orchestrated collaborations that drove the creation and documentation of the system-near export.
For example, a big part of our PVS export was working with the developers to document and debug the generation of XML files.
In the case of Coq, Claudio Sacerdoti Coen's kernel export plugin \cite{coqtoxml_csc} took several person-months and was developed specifically for the occasion.
The Isabelle export included some 
 person-months of work by Makarius Wenzel to invest heavily into the PIDE infrastructure to make the export possible.

\subsection{Compositionality and Trustworthiness}

An important question is whether a library export is correct, particularly in the realistic scenario where we can translate a library to another system only without proofs (because translating proofs is generally much harder than translating everything else).
Moreover, even if all proofs can be exported and rechecked (both of which are problematic, see below), there is no guarantee that they were exported correctly.
In fact, typical exports go through multiple bugs where proofs are falsely reported as successfully rechecked.
For example, even a bug as blatant as accidentally generating all theorem files as empty strings (which then trivially recheck) can be surprisingly hard to notice because these exports involve large sets of large files often containing formalizations from domains that neither the prover expert nor the \mmt expert understand.

Rarely used or experimental language features pose a related problem.
Often the prover expert notices missing cases in her export only when the feature is actually used in a library, which may be very far down the line.
For example, in our Coq export, we encountered kernel features that were not used at all in any of the libraries that we could build with the latest Coq version.
We noticed that issue because we actually looked for uses of the feature in order to reverse-engineer the intended semantics from the way it was used.

To ensure the correctness of exporters, we found the skill-separating export approach above extremely valuable.
It allows for the system-near export to be relatively trustworthy as it is a straightforward serialization of the kernel data structures, and it pushes all logical transformations (e.g., eliminating subtyping) to a later phase.
Mixing the serialization of the data structures with such logical transformations is not only undesirable (because information is lost) but has also proved error-prone because the former is a major implementation effort and the latter extremely difficulty theoretically.

Even so, we found that any logical transformation in later phases must remain simple and compositional: any non-compositional transformation is usually a huge theoretical challenge, ends up incomplete or hacky because the theory did not cover all cases that are actually used in the prover, and is the first thing to break upon new prover releases.

Thus, we have opted to avoid non-compositional transformations such as eliminating PVS subtyping, Coq universe constraints, or Mizar structures.
If in doubt, we mimicked the language feature in \mmt by extending the logical framework.
Thus, any feature eliminations (which may be necessary, e.g., to translate into a less expressive language) are optional and can be delayed until needed.

\subsection{Generating the Prover-Near Export}

There are multiple general approaches to \textbf{setting up the export} of libraries.
The most direct approach of exporting from the source files basically never works in practice because it would essentially require reimplementing the entire prover.
Instead, only exports that use the kernel data structures are feasible.
\emph{Instrumenting} the kernel means that the library is checked by the kernel, and kernel hooks are used to generate the export.
It requires rechecking the entire library. (If proof exports are not needed, it may be possible to tweak the kernel to skip all proofs, which is the most expensive part.)
But if the kernel is small and well-structured, instrumentation is relatively easy to implement and maintain.

Alternatively, we can use already existing kernel-generated files (e.g., binary files for Coq or XML files for Mizar).
This requires no modifications of the kernel and is faster.
If these kernel-generated files are already used internally (as for Coq and Mizar), maintenance is assured and thus bit-rot avoided.
But this approach is limited to the information that the kernel generated, which is often not exactly what one is interested in for the export.
And the file structure may be subject to change, thus breaking any tool that picks up on them.
For that reason, we instrumented the Coq kernel even though it already generates binary files.

If the kernel-generated files are not used internally but are only written out for export purposes (as for PVS), it is easier to negotiate changes to their format (e.g., to include additional information).
But there is little assurance that they were generated correctly in the first place.
In that case, an initial debug cycle may be necessary until the files are generated correctly.
But that is still much easier than building an instrumentation from scratch.

A subtle practical issue we encountered with kernel-generated files in multiple situations (e.g., PVS/NASA) is a reliance on case-sensitive file systems.
Because logical names declared inside source files are usually case-sensitive, care is needed when generating a separate file for every such declaration: developers must encode the names to make the generated file names unique without relying on case.
Otherwise, exports may be entirely unusable on some systems, e.g., when cloning a git repository on a case-insensitive file system, the second file overrides the first; then git detects a change, and every subsequent pull fails.

\subsection{User-Level vs. Kernel-Level Syntax}\label{sec:elab}

The main drawback of kernel-based exports is that the \emph{user-level} data structures in the source can be very different from the \emph{kernel-level} data structures that make up the export.
Most systems employ complex processing chains that \emph{elaborate} the user-level into kernel-level data structures.
This includes beneficial steps such as disambiguating identifiers and notations, inferring omitted information (types, implicit arguments, proof steps, etc.), and computing normal forms.
But this information can also eliminate high-level structure such as module system (Coq records, Mizar structures), processing hints (Coq's unification hints), or constructs that are internally defined as abbreviations (e.g., Isabelle's inductive types, Mizar's implications).

Thus, a lot of high-level structure is elaborated away on its way to the kernel.
But it would be extremely valuable for virtually any application other than proof-checking, and it is extremely difficult or even impossible to recover this information in an export.
For example, most systems provide some support for common high-level constructs such as mathematical structures (modules, records, etc.) and term languages (inductive, co-inductive types), and when translating between systems or integrating libraries, it is highly desirable to match these up with each other.
But in the common situation where different systems elaborate these high-level constructs in different ways, this is not possible.
Exports that preserve this user-level structure remain an open problem for all systems.
For example, we only had the resources to preserve Coq sections and Isabelle locales but none of the many other high-level constructs.

The representation of these user-level features in logical frameworks is much harder and less well-studied that the representations of kernel-level  languages.
That is partially because kernel-level syntax is often defined using the highly standardized methods of context-free grammars and context-sensitive type and proof systems, which yields a strong relation between specification and implementation.
The elaboration of user-level syntax, however, often uses relatively unconstrained programs in Turing-complete programming languages.
To catch and mimic this practice, \mmt provides two mechanisms for representing elaboration-based features.
Declaration patterns \cite{HKR:extending:12,Iancu:phd} are used for statements whose elaboration can be defined declaratively by simple rules.
We use it to define HOL type definitions and the various definition principles of Mizar.
Secondly and more generally, arbitrary elaboration can be defined in the Scala language underlying \mmt.
We use this to define a few advanced features such as PVS includes and Coq sections, and we could also use it to capture Isabelle's heavily-used corresponding mechanism for arbitrary elaboration.
We recently described details and collected more examples in \cite{MueRabRot:rslffml20}.

A related problem is the destruction of \textbf{structure-sharing}.
Many systems internally employ sophisticated structure-sharing to reduce memory requirements.
Naive exports often cannot reuse this structure-sharing easily, potentially leading to enormous blowups in the generated intermediate files and ultimately in the export.
For example, Isabelle exports of the entire AFP that include all proof terms have not been done so far at all because of this explosion.
New research into structure-sharing-preserving exports is needed to make proof exports for Isabelle possible.
Apart from the practical difficulties, even the theoretical design of such exports is more difficult than it may appear at first.
For example, Coq's structure sharing is syntactic rather than semantic, i.e., Coq shares two terms with free occurrences of variables even if they refer to different variable declarations.
It is not obvious how to treat this best in an export.

\subsection{Exporting Proof Objects}\label{sec:proofexp}

The conflict between user-level and kernel-level is particularly critical for the export of \textbf{proofs}.
It is still unclear what the best way to export proofs is.

The export of \emph{kernel-level proofs} is often straightforward, but the proof objects become huge.
This is particularly severe when proofs include automatically found parts, which may be much larger than necessary.
For example, our Coq export runs out of memory on some very large proofs in the Feit-Thompson proof \cite{oddorder}.
Claudio Sacerdoti Coen speculates this might be because an unnecessarily large proof term is built internally.
But a detailed investigation has so far been impossible because it would require precious system export resources.
Moreover, kernel-level proofs have only limited value, independent proof checking essentially being their only use.

The \emph{user-level proofs} are much more interesting for, e.g., viewing, searching, reusing, or translating libraries, but they can usually not be exported or only be exported in source-near syntax (strings in the worst case) that lack the information inferred by the prover.

A third option is to export \emph{dependency-only} proofs, where a special object $\mathit{dependsOn}(\mathit{listOfIdentifiers})$ is used as the proof that records only the axioms and theorems that were used.
This is sufficient for some applications such as change management and premise selection as done in \cite{holyhammer}.

In any case, kernel-level proofs may still have logical gaps making them insufficient even for proof checking, e.g., if automated proving or decision procedures are part of the kernel (as for Mizar and PVS).
They may also have pragmatic gaps if the logic includes powerful computation (as for Coq) as opposed to logics that record computations in explicit rewrite steps (as for Isabelle).
Depending on the application, these gaps may be seen as advantages (skipping low-level steps that can be more easily recreated after translating the proof to another system) or disadvantages (precluding the rechecking of the proofs).

In general, a major lesson of our exports is that the community has not yet converged on a good solution for exporting mid-level proofs that combines the relevant structure of user-level proofs with the inferred information and re-checkability of kernel-level proofs.
In particular, it would be great if applications processing proofs could gradually choose the level at which a subproof is seen.
For example, a proof translation tool should try to map every high-level proof rule to the target's counterpart whenever such a counterpart exists.

Therefore, all of our exports have de-emphasized proof export and usually opted for dependency-only exports.
We only experimented with a few kernel-level proofs for Coq to test the scalability of the approach.

\subsection{Heterogeneity}

Most proof assistants use what we call the \emph{homogeneous} method, which fixes one foundational logic and builds on it using conservative extensions such as definitions and theorems.
Thus, all domain knowledge is ultimately represented in terms of the same primitive concepts.

Advanced user-level declarations such as inductive type or recursive functions definitions are typically justified by meta-theorems that establish their conservativity.
The meta-proof may be used explicitly in the system by elaborating user-level declarations (as in Isabelle) or be left to the literature about the system (as when adding inductive types to the calculus of constructions in Coq).
Both can be problematic: The former can lead to more numerous and less intuitive declarations in the kernel, which can cause a strong disconnect between user and kernel-level declarations.
The latter may cause very subtle issues (in fact so subtle that discussions usually stay limited to small circles of experts) when the meta-theorems do not perfectly match the implementation, e.g., if the conservativity of two features is proved individually relative to the base calculus without noticing that their combination in the implementation may violate conservativity.
As described in Section~\ref{sec:elab}, the former of these two is most critical from the perspective of exporting and obtaining system interoperability.

We speak of the \emph{heterogeneous} method if systems define theories that encapsulate choices of primitive concepts or notions and then considers truth relative to a theory.
The homogeneous method can be seen as the special case where a single theory is fixed, e.g. the underlying set theory of Mizar or the theory of an infinite type in HOL Light.

The heterogeneous method allows capturing the mathematical maxim of stating every result in the weakest possible theory and moving results between theories in a truth-preserving way. (The formal tool to capture this moving operation are theory morphisms.)
This is often called the \textbf{little theories approach} \cite{littletheories} as pioneered by IMPS and supported by Isabelle and PVS.
Even if heterogeneous reasoning is possible, it may be \textbf{optional}, e.g., users of Coq and Isabelle often do not make use of heterogeneity even when they could.
PVS and IMPS on the other hand force users to state results inside theories, and IMPS strongly encourages the use of theory morphisms.

We speak of \textbf{external} heterogeneity if it is captured by an explicit language feature that sits on top of the base logic, such as Coq modules, Isabelle locales, or PVS and IMPS theories.
External heterogeneity is particularly valuable for exporting and system interoperability: it allows retaining the heterogeneous structures in \mmt and then matching it to the corresponding construct in other systems.
This is critical as many theories are much weaker than the base logic they are written in:
Translations between different base logics are often prohibitively difficult, whereas translations between corresponding theories can be much simpler.
For example, translating from calculus of constructions to higher-order logic is very hard, but translating from the theory of natural numbers in the former to the theory of natural numbers in the latter is much easier.

Therefore, from the perspective of library integration, it is critical to use heterogeneity and to preserve it in exports.
We were able to directly preserve it for PVS and IMPS.
For Isabelle, heterogeneity only exists at the user-level, and elaboration reduces it to a homogeneous kernel-level formalization.
But we were able to reconstruct the user-level locale structure and export it in addition.
We also preserved it for Coq, but this has limited practical value as Coq modules are not widely used throughout the libraries.

We speak of \textbf{internal} heterogeneity if it is captured by record types (or a similar language feature like Mizar structures) within the base logic.
Record types are supported by most proof assistants and are the predominant source of heterogeneity in Coq (most prominently used in the Mathematical Components project \cite{gonthier_packaging}).
We discussed this issue in more detail in \cite{MueRabKoh:tat18}.
Internal heterogeneity provides many of the same and some additional advantages as external heterogeneity.
Its main drawback arises in library integration:
As record types are just types in the base logic, it can be very difficult to spot which of them correspond to external heterogeneity features in other libraries, e.g., to automatically match up Coq records with corresponding Isabelle locales.
On the contrary, matching up Isabelle locales with PVS theories, while also difficult, is much easier.
In any case, either remains future work.

\subsection{Toplevel Binding}

Most provers allow for what we call \emph{implicit toplevel binding}.
These are free or specially bound variables that are treated as if they were universally quantified at the beginning of the containing statement.
Examples are type variables in HOL, type and subtype variables in PVS, and function and predicate variables in Mizar.
Internally, universes in Coq behave accordingly although no user-level syntax is provided.

Characteristically, the respective logic does not feature the corresponding universal quantifier so that these variables can never be bound in subexpressions.
Their implicit binding at the toplevel of a statement is the only way to introduce them.
Technically, allowing such toplevel binding changes the logic.
However, as a general rule, this is conservative over the base logic: top-level binding can be eliminated in favor of all (i.e., usually infinitely many) ground substitution instances.
This is well-known from axiom schemas in axiomatic set theory: in order to stay within first-order logic, authors often state a single second-order axiom as a set of infinitely many first-order substitution instances of a schema.

While this is a relatively minor theoretical observation, we mention it here explicitly because it is not widely known in all its generality.
Therefore, the details in individual provers are not always cleanly documented or implemented, and their treatment can occasionally be confusing when writing exports.

Interestingly, when defining the logics in an LF-like logical framework, the formalization of these toplevel binders requires no additional work: they are directly represented using the $\Pi$-binder of LF.
This can be seen as a rigorous argument for why these binders are in fact conservative.

\subsection{Non-Logical Information}

There are a number semantically irrelevant language features that are often critical to export as well.
In some cases, it is even strongly advisable to prioritize them higher than some logical features.

By far the most important non-logical aspect of an export is attaching \textbf{source references}.
These annotate each element in the export with the corresponding physical location (if any) in the source (i.e., file, line, column).
Even though some form of source references is needed anyway for provers to report useful error messages, it is not always easy to preserve them in the export, depending on how they are processed and stored internally.
This can be especially difficult if elaboration substantially changes the syntax tree.
But source references are crucial to enable many applications of the exports because they allow pointing users to the human-readable sources whenever they interact with a part of the export, especially in the typical case where the kernel-level data structures in the export may look entirely unfamiliar to the user.
A typical example is search, where one wants to search through the export but show results in the source.
Source references should ideally be present \emph{everywhere}, i.e., for each subexpression, but are at least needed for every \emph{statement} (definition, theorem etc.).
In the worst case, source references for statements can be \emph{recoverable} by parsing the source in a separate process.

Another  obviously desirable non-logical feature are \textbf{comments}.
Exporting comments is relatively straightforward iff they are preserved during elaboration.
\mmt can even represent alternated nesting of formal and informal text.
But elaboration often does not preserve comments --- a common approach to build systems is to discard comments very early, even during parsing.
Isabelle is special in that it includes structured markdown-like syntax for informal text at the same level as formal developments.
These are easy to preserve in the export.

Finally, semantic web-style \textbf{ontological abstractions} have proved very successful in many areas such as biology and medicine.
The analogues for proof assistant libraries systematically abstract away all symbolic expressions (types, formulas, proofs, etc.) and only retain identifiers (of modules, statements, etc.) and properties and relations on them such as authorship, dependency, or check time.
While these do not capture the entire semantics of the library, they are extremely powerful for limited problems:
Their level of abstraction is sufficient for shallow services such as semantic navigation, search, or querying.
Even better, at this level, the huge formal differences between the libraries disappear and services can easily span multiple libraries, e.g., to find related formalizations in different libraries.
Moreover, standardized formalisms such as RDF \cite{rdf} and SPARQL~\cite{sparql} and highly scalable tools are readily available.
While such applications were already envisioned years ago, e.g., in \cite{AGSTZ:ContMathSearchWhelp04,aspinall_queries}, they could so far not be realized at large scales because provers were unable to export the necessary data.
Therefore, we have recently carried out two major case studies in supplementing the exports described so far with exports of ontological abstractions: in \cite{CKMRSW:ulo:19}, we define an upper library ontology and use it to export linked data representations for the Isabelle (40M RDF triples) and Coq (12M RDF triples) libraries.
This enables, for example, querying for all Isabelle or Coq theorems about a given mathematical concept.
Corresponding extensions of the other exports are straightforward conceptually; they still require a considerable investment in collaboration with the system expert but are significantly easier than full logical exports.

\subsection{Library Structure}

Due to the tremendous logistical challenges in maintaining large libraries with a many authors over decades, multiple models for the \textbf{structure of libraries} have been developed.
Typically each system is \emph{integrated} with a standard library that is co-released with the system, often maintained in the same repository as the system's source code.
This library can be comprehensive or small and be extended by any number of external \emph{distributed} libraries ---
HOL Light with a single integrated library and Coq with many distributed libraries can be seen as opposite extremes on this spectrum.

For external libraries, a critical question is whether these are \emph{co-maintained} with the system, usually as part of the regression test suite that is checked before every system release.
In that case, the library can be a collection of independent user-\emph{submissions} or a single curated \emph{coherent} body ---
Isabelle/AFP with many relatively unrelated submissions and the Mizar library carefully curated by a small committee are opposite extremes on this spectrum.
Both are officially co-maintained with the system.
PVS/NASA is special in that it is a single external library that is developed independently from the prover; in practice, it is used as the main regression suite for the system.

Notably, if many distributed libraries exist (as for Coq) or if the central library is submission-based (as for Isabelle), there is no guarantee that all developments are compatible with each other.
They may depend on different system settings (e.g., impredicativity for Coq) or system states (e.g., loaded language extensions or automations), or try to register the same names.
Therefore, it is not always possible to speak of \emph{the} system export; even within a single library, different parts may be inconsistent with each other.

Table \ref{tab:general-aspects} gives an overview over the general aspects of the six exports we will discuss below.

\begin{table}[ht]\centering
\begin{tabular}{|l|lll|}\hline
      & Coq & HOL Light & Isabelle \\\hline
 Foundation & type th. & HOL & HOL \\
 Lib. org. & distr. & integr. & subm.-based \\
 Libs co-released & - & + & + \\
 Kernel-generated & binary & - & XML \\
 Export setup & instr. & instr. & instr. \\
 Proof export & low-level & low-level & low-level \\
 not in proofs & computation & - & - \\ 
 heterogeneous & int., opt. & no & ext., opt\\ 
 th. morph.    & - & -& + \\
 Src refs & statements & recoverable & everywhere \\
 \# \mmt decls. & $170k$ & $20k$ & $1,500k$ \\
  \# RDF triples & $12M$ & N/A & $40M$ \\\hline\multicolumn{4}{}{}\\\hline
  & IMPS & Mizar & PVS \\
 Foundation & HOL+X & set th. & HOL + X\\
 Lib. org. & integr. & curated & distr., co-m. \\
 Libs co-released & + & + & + \\
 Kernel-generated & JSON & XML & XML \\
 Export setup & from JSON & from XML & from XML \\
 Proof export & low-level & high-level & high-level \\
 not in proofs & - & automation & automation \\ 
 heterogeneous & ext. & no & ext. \\ 
 th. morph.    & + & - & + \\
 Src refs & recoverable & none & everywhere \\
 \# \mmt decls. & $2k$ & $70k$ & $25k$\\
 \# RDF triples & N/A & N/A & N/A\\
\hline
\end{tabular}
\caption{General Aspects of Theorem prover exports}\label{tab:general-aspects}
\end{table}



\section{The Coq Libraries}\label{sec:coq}
 Coq is based on constructive type theory and offers strong support for general purpose (pure) programming and theorem proving.
It supports a diverse user community large enough to require decentralized library maintenance and hosts multiple flagship projects from mathematics (such as the Odd Order proof \cite{oddorder}) and computer science (such as the CompCert C compiler \cite{compcert}).

Our Coq export was carried out together with Dennis M\"uller and Claudio Sacerdoti Coen.
The details were published in \cite{coqtoxml_csc,MRS:coq:19}.

\subsection{Language}

\paragraph{Overview and Formalization}
Coq is based on the calculus of inductive constructions (CIC), which can be roughly seen as dependent type theory plus universe hierarchy plus (co)inductive types.
The exact status of propositions is subtle, but essentially every proposition is a type.
Generally, the most desirable representation in a logical framework is a Church-style typed one: it uses $\tp: \type$ to represent every Coq-type $A$ as an LF-term $\enc{A}:\tp$, and $\tm: \tp\to\type$ to represent every Coq-term $t:A$ as an LF-term $\enc{t}:\tm\,\enc{A}$.
This has the advantage that only well-typed terms can be encoded at all, and Coq's typing rules are captured by the LF type system.

But this representation requires more type annotations than typically present.
For example, if $A\Rightarrow B$ encodes simple functions, the application operator is represented as
$\apply: \Pi_{A,B:\tp}\tm (A\Rightarrow B) \to \tm\,A\to\tm\,B$.
Thus, a Coq application is represented as an LF term that records its input type $A$ and output type $B$. 
(Actually, Coq uses a dependent function space, but the same argument applies.)
In most cases, this additional information is advantageous and is already present in the kernel data structures or can be inferred relatively easily, by calling either Coq functions during the export or logical framework functions in \mmt.
But it significantly increases the size of the export.
This size explosion problem can be alleviated partially by using rewriting as in \cite{coqine}.
This allows annotating only $\Pi$-expressions with types but not applications.

In the case of Coq, we exported low-level proof terms, and a typed representation would not have been feasible yet.
We used an untyped Curry-style representation based on a single type $\expr:\type$ of all Coq terms and types, and a separate typing judgment $\of:\expr\to\expr\to\type$ of typing derivations.
Now the application operator is simply formalized as $\apply:\expr\to\expr\to\expr$ together with appropriate typing rules.
Even this untyped encoding ran out-of-memory on some of the largest proofs in the Feit-Thompson proof \cite{oddorder}.

The untyped encoding also loses the types of Coq-identifiers $c:A$, which simply become $c:\expr$ in LF.
To remedy this, one can represent $c:A$ using two declarations $c:\expr,\,c\_type:\of\,c\,\enc{A}$.
To avoid making the encoding non-compositional in this way, we used \mmt's ability to flexibly extend the logical framework: We switched to LF with predicate subtypes, and defined $\tm\,A$ as the subtype of $\expr$ containing only those $e:\expr$ that satisfy $\of\,e\,A$.
This allows compositionally representing $c:A$ as $c:\tm\,\enc{A}$ as in Church-style encodings.

\paragraph{Universes}
Coq's universe hierarchy makes the Church representation even more complex because a second parameter is needed: $\univ:\type,\;\tp:\univ\to\type,\;\tm:\Pi_{U:\univ}.\tp\,U\to \type$.
The application operator now takes four additional arguments instead of two as in
$\apply: \Pi_{U,V:\univ,A:\tp\,U,\Pi B:\tp\;V}\tm\, M\,(A\Rightarrow B) \to \tm\,U\,A\to\tm\,N\,B$.
Moreover, additional formalization steps are needed to compute the universes $M$ and $N$ from $U$ and $V$ according to the Coq typing rules.
Finally, as Coq's universe hierarchy is cumulative, the Church representation breaks down entirely as injection functions have to be inserted non-compositionally all over the place to cast types into higher universes.
This is an additional reason why only a Curry representation is feasible at this point.

Coq source syntax only contains the identifier $\texttt{TYPE}$ without clarifying its universe, and the system internally introduces a fresh universe identifier for which it then infers constraints.
Because these constraints are affected by how an identifier is declared and how it is used, this information must be maintained globally.
If an identifier is used in multiple different source files or even libraries, the resulting constraints may be inconsistent with each other.
Technically, a Coq export must include these universe constraints and treat them as assumptions relative to which the theorems are stated.
This is what we did.
But as this information is usually not interesting to the user and not portable to other provers, it remains an open question how to handle these constraints at all.

\paragraph{Inductive Types}
The above description is only accurate if we ignore Coq's (co)in\-duc\-tive types.
These are very complex, e.g., using multiple kinds of parameters and offering primitive operators for recursion and pattern-matching.
Capturing the associated typing rules in a declarative logical framework has so far been out of reach, be it Church or Curry-style.
And eliminating them is (even if we are willing to lose this high-level feature) is a prime example of a theoretically possible but practically doomed non-compositional logical transformation.
Therefore, our representation had to go outside LF by using untyped \mmt symbols to represent recursion and pattern-matching.
Since our representation was untyped anyway, this allowed exporting all Coq expressions regarding (co)inductive terms without significant additional loss.
A representation in a stricter framework like LF or Dedukti ($\hat=$ LF modulo) would be very difficult.

\subsection{System}

The Coq system has grown for several decades, and its internal workings are extremely complex.
More recent reimplementations of essentially the same language like Matita \cite{matita} or Lean \cite{lean} have been able to simplify the implementation drastically.
But because they are neither binary nor source compatible with Coq (Matita could read Coq binaries at one point but then diverged), they cannot process the huge Coq libraries.

Recently, Coq development has increasingly focused on giving users more control how their input is interpreted before it reaches the kernel.
Type classes and canonical structures/unification hints are the most important examples.
These are not visible to the kernel and therefore not part of exports like ours.

Similarly, all Coq proof objects are low-level $\lambda$-terms, and high-level structure via tactic languages is elaborated away.
This makes Coq exports very big if they include proofs.
Because computation (in particular, recursive functions on inductive types) is a kernel feature and thus not part of the proof terms at all, some complex proof steps do not have an effect on proof size.
Therefore, it was feasible to export all proof terms.

The Coq module system is primitive and fully visible in the kernel.
That allows fully preserving the modular structure.
However, the trend in Coq development is towards using record types (which are treated as inductive types with one constructor) instead of modules.
While visible to the kernel, this modular structure is very hard to recover.
Therefore, there is no export yet that can identify this modular structure.

Maybe surprisingly, the Coq system includes several imperative aspects.
For example, a section is checked as usual, but when closing a section the internal state of Coq is rolled-back imperatively and the section is installed in the way in which it is visible from the outside.
That makes sections impossible to export via the kernel-generated binary files and difficult to export via kernel instrumentation.
We opted for the latter.

\subsection{Libraries}

Coq has reached the usage size, where the central maintenance of libraries is no longer feasible.
Instead, the Coq library has been factored into hundreds of repositories with a somewhat standardized build process.
This allows distributed maintenance of the library.
But it also means that not every repository always builds with the latest version of Coq.
For example, when we ran our export in early 2019, only around 70 of around 250 repositories could be built, including the MathComp libraries (the situation has improved since then).

We used the XML files produced by kernel-instrumentation in \cite{coqtoxml_csc} for our export \cite{MRS:coq:19}.
Due to the distribution of libraries, significant additional scripting was needed to detect all libraries, identify their metadata and dependencies, and iterate through them.
The metadata does not include which toplevel Coq namespace(s) a library's declarations are in so that additional checks were needed to determine which declarations to export.

Alternative exports are presented in \cite{assafphd,coqine} using Church representations in Dedukti after eliminating (co)inductive types and in \cite{coqhammer} to first-order logic for the purpose of premise selection, but none covers the entire language.

\subsection{Outlook and Open Challenges}

The biggest future challenge for Coq is to \textbf{scale up} the export.
Our export is deceptively scalable --- to the point of including all proof terms --- because it uses an untyped representation in the logical framework.
Switching the export code to work relative to a typed representation would be straightforward but would yield much bigger exports.
The usefulness of the untyped representations is limited because any kind of type inference of re-verification must implement the Coq type system from scratch.
The rewriting-based representations in Dedukti may alleviate this problem.
But we expect that only a systematic solution to proof export as indicated in Section~\ref{sec:proofexp} will eventually be successful.

This will require two investments:
Firstly, Coq must process its proofs in a way that allows recovering \textbf{mid-level proof terms} --- terms that includes more information than the users tactic script (e.g., all intermediate proof states and tactic invocations) but less than the low-level $\lambda$-terms.
Secondly, proofs that are the result of search and other computations must be refactored to ensure they are not massively larger than necessary.
That would also help cause of program extraction --- it is plausible that some programmatically found current Coq proofs are so convoluted that they would not allow for extracting elegant programs.
As a simple immediate step, it may help for Coq to display the internal size and possibly structure of every proof.
That would allow users to notice when proofs are much bigger than expected.

The second huge challenge, which is also a requirement for a \textbf{typed representation}, is to define a declarative representation of the Coq typing rules in a logical framework.
Here, the treatment of (co-)inductive types and recursion is currently not possible with state of the art frameworks.
We circumvent this issue by using untyped representations; the exports in Dedukti circumvent it by using a typed representation that does not cover these features and applying a non-compositional transformation to eliminate them during the export.
We expect building appropriate frameworks and applying them to Coq will be an effort measurable in person years (even when using \mmt for rapid prototyping).
This would ultimately also require sorting out a few subtleties in the inner workings of Coq such as the treatment of record projections (which are primitive in the kernel for efficiency even though records in general are not) or the various corner case in the handling of inductive types.

A major Coq-specific logistical challenge is given by the \textbf{distributed library}.
While necessary due to the size of the user community, it brings a host of new maintenance problems such as different libraries depending on different versions of Coq.
It is likely that this problem will not go away as new versions of Coq will be released faster than all existing content will be adapted.
Therefore, new tools will be needed to manage the symptoms.
However, this is a general issue facing the Coq community, and we expect good package and build managers to emerge in the near future, which can then be integrated with exports.
That will make the scripts obsolete that we used to run our export over all libraries and will ensure the maximal number of libraries can be exported.

Finally, while large scale library integration and system interoperability is mostly a problem of the future, we can already foresee that the predominant use of record types for heterogeneity will cause problems.
A good compromise might be to annotate those record types that are meant to be used in the sense of mathematical theories so that exports can translate those records differently.
This might require limiting the use of such records: for example, record types cannot be mapped to Isabelle locales if they occur as the types of non-toplevel bound variables.



\section{The HOL Light Library}\label{sec:hollight}
 HOL Light is a minimalist implementation of standard higher-order logic without complex additional features with a small and easy to understand kernel.
It is maintained essentially by a single developer and features a large integrated and coherent library.
That made it flexible and scalable enough to be chosen for the Flyspeck project \cite{flyspeck}.
Due to its simplicity, the language can be embedded into most other logics (Mizar, which lacks $\lambda$-abstraction being the main counter-example).
This combination makes it very popular source language for translations \cite{hol_nuprl,hol_coq,holide,hol_isahol_kaliszyk}, and it was the first for which a major library translation was done \cite{hol_isahol}.

Our HOL Light export was carried out together with Cezary Kaliszyk.
The details were published in \cite{KR:hollight:14}.

\subsection{Language}

We formalized the HOL Light in LF using the analogue of the Church encoding sketched in the section on Coq.
Because this was the first export we did, we refrained from exporting proof terms at all so that the Church encoding did not present a scalability hazard.
Instead, we exported dependency-only proofs.

\subsection{System}

HOL Light is implemented inside the OCaml toplevel with only a few tweaks for parsing expressions.
This makes the kernel very easy to instrument for generating an export.
However, it makes it near impossible to export any of the high-level features used to build the library as these are arbitrary OCaml programs on top of the kernel.
For example, the kernel is not even aware of the list of theorems, which are instead maintained by the OCaml context.
Any high-level or tactic-based proofs are invisible to the kernel and only low-level exports scale to the whole library.
This is particularly unfortunate in the case of HOL Light because the tactic language is highly imperative, and even the source-level proofs can be very hard to read.
To remedy this problem, \cite{hollight_hiproofs} patches individual tactics in order to export tactic applications in addition to kernel-level.

Because the source files can contain arbitrary OCaml code, recovering source references is difficult in principle.
However, because the library uses consistent conventions it is relative easy to write parsers that identity the source lines of statements.

HOL Light inherits the OCaml module system.
However, this is rarely used in practice.
In any case, any modular features are invisible to the kernel.

\subsection{Library}

The HOL Light library is highly integrated with the system and maintained by the same single person.
That makes it very smooth to export by kernel instrumentation.
Our export \cite{KR:hollight:14} modified the existing instrumentation to output OMDoc files directly.
Additionally, it accesses an internal table with notations in order to preserve those in the export.

The most important other library is the Flyspeck proof \cite{flyspeck}.
We have not exported it because it is extremely big. 
It would also not be particularly interesting (except for porting the proof itself) as most generally reusable formalizations have been migrated from Flyspeck to the main library already.

\subsection{Outlook and Open Challenges}

Due to the simplicity and stability of the language and the system as well as the coherence of the library, HOL Light is among the easiest systems to export and to maintain an export for.

But its low-level kernel that users can feed arbitrarily with programmatically generated proofs makes it extremely difficult to export anything but low-level proof terms.
Therefore, we cannot expect naive proof terms exports of large projects such as Flyspeck to scale without systematic changes.
Even if a standard for mid-level proof terms in the sense of Section~\ref{sec:proofexp} exists, it will be difficult to \textbf{export proof terms} from HOL Light.
The most realistic option would be to instrument not only the kernel but also individual tactics, and efforts in this direction have been made already \cite{hollight_hiproofs}.
But as users can write arbitrary tactics, this will require a significant investment of expert knowledge to be comprehensive.
Establishing best practices and special macros for tactics with an eye towards instrumentation is helpful here.

Another major challenge is that HOL Light uses almost exclusively the homogeneous method, which makes it more difficult to integrate with other libraries.
Here it would be helpful to research the \textbf{automated introduction of heterogeneity} by abstracting from the assumptions theorems and grouping them according to their assumptions.



\section{The IMPS Library}\label{sec:imps}
 IMPS was one of the earliest proof assistants.
It used a higher-order logic extended with features for subtyping and partial functions.
Despite pioneering some original features that are still of interest for proof assistants today, it has fallen out of use before building a large library.
Today it only runs on two machines: one installation by one of the original developers (Farmer), and one of our's for our export.
Its inclusion in our set of case studies was motivated not by size of library and user base (as for the other provers).
Instead, it was motivated by (i) IMPS's commitment to the heterogeneous method, which will deserve more attention to make future library integrations more feasible, and (ii) to permanently archive its library before the technology to process it dies.

Our IMPS export was carried out together with Bill Farmer and Jones Betzendahl.
The details were published in \cite{imps_oaf}.

\subsection{Language}

IMPS's underlying logic, called LUTINS, is a variant of Andrews' higher-order logic $\mathbf{Q}^0$~\cite{andr:truth86}.
Most characteristically, all functions are (potentially) partial, and terms may be undefined.
The latter is captured by a primitive unary predicate for definedness.

IMPS uses a limited subtyping system: underneath each base type, a hierarchy of subtypes (called sorts in IMPS) may be introduced.
Due to the partiality, the domain type of a function type is only an upper limit on the set of arguments for which the function is defined.
Interestingly, that makes the function type operator covariant in both arguments.

Base types for individuals and binary booleans is built-in, and undefined constant applications that return booleans are considered false.
Other base types and their subtypes can be declared in theories.
Contrary to most other proof assistants, IMPS systematically uses the heterogeneous method, and theories routinely contain base types and operations on them whose properties are not given by definitions but by axiomatizations.

Some flexibility exists in how to represent LUTINS in LF.
It is straightforward to give a Curry-style representation using $\tp:\type$, $\tm:\type$, and $\of:\tm\to\tp\to\type$.
Alternatively, we can give a Church-style representation for the maximal types and then a Curry-style representation for the sorts underneath each type: this would use $\tp:\type$, $\tm:\tp\to\type$, $\sort:\tp\to\type$, and $\of:\Pi_{A:\tp}\tm\,A\to\sort\,A\to\type$.
The latter allows LF to perform some type-checking automatically at the cost of making the representation a bit clunkier.
For example, the latter requires all polymorphic operators to take two arguments: one for the type $A$ and one for the sort $S:\sort\, A$.
Finally, we can use a complete Church-style representation using $\tp:\type$, $\sort:\tp\to\type$, and  $\tm:\Pi_{A:\tp}\sort\,A\to\type$.
This yields the most elegant representation of the typing rules, but it requires explicit casts whenever subsorting is applied.
It is relatively easy to switch between these, and the details of the trade off are ongoing work.

The only subtlety is the handling of binders as bound variables always range over defined values:
All typing rules that introduce variables ($\lambda$-abstraction, universal introduction, existential elimination) must add assumptions to the context that assume the definedness of (the instances of) the bound variable.
Correspondingly, the rules that perform substitutions ($\beta$-reduction, universal elimination, existential introduction) must have premises that ensure the substituting term is defined.
In the first two alternative representations, this can be taken care of by formulating the rules in such a way that $\of\,t\,A$ only holds if $t$ is indeed defined.

\subsection{System}

IMPS is implemented in LISP, and uses a LISP-like source syntax.
Because it has not been maintained for years, it is rather difficult to work with.
But previous work \cite{imps_omdoc} had already developed a partial instrumentation to generate \omdoc files.
Because that work predates \mmt and did not use any rigorous semantics for \omdoc, let alone logical frameworks, it is better seen as system-near XML export rather than as a semantics-preserving export.

We adapted and extended this earlier work by developing a system-near export in JSON.
Because the IMPS parser does not record comments or source references, we recovered source references by writing a fresh parser for the IMPS source syntax (in Scala) solely for the purpose of recording this non-logical information.
Eventually, we wrote a Scala program to merge the two data structures to build \mmt data structures in Scala, which could be easily serialized.

IMPS stores proof objects that record tactic invocations made through a graphical interface.
In addition to user-level tactics, special tactics (called Macetes in IMPS) are automatically generated from theorems of certain shapes, e.g., to turn theorems in Horn form into proof rules.
This is particularly important to automate the tedious reasoning about subtyping and definedness.
These proof objects contain gaps where IMPS automation was used.
Thus, the proofs are already roughly mid-level proofs that can be more easily translated to other provers, and we expect that the missing steps can be recreated by the automation in those provers relatively easily.

\subsection{Library}

Because the library is small, frozen, and part of the IMPS source code, it is relatively easy to export it once and for all in a single run.

Only two subtleties arose in the structuring mechanisms used by IMPS.
Firstly, IMPS theories are split into languages and theories and organized via theory ensembles.
But that idiosyncrasy is arguably less critical to preserve, and we represented all as \mmt theories.

Secondly, IMPS's treatment of theory morphisms includes a subtle imperative part: after defining a theory $T$ and some theory morphism $m:S\to T$, users can install the morphism by imperatively adding the translations of $S$-theorems to $T$.
While straightforward in an implementation that imperatively maintains theories as tables of declarations, it is problematic in exports.
We could mimic it \mmt only by defining a new theory $T_m$ that conservatively extends $T$ with the corresponding definitions.

Finally, for a few rarely used unusual features such as the definition of what IMPS calls quasi-constructors, we simply collected all instances and formalized them in LF manually.

\subsection{Outlook and Open Challenges}

Our export provides an archive of the IMPS library in \omdoc.
Additionally, because we also built system-near JSON representations and a fresh parser of the IMPS library, other exports can be built off these IMPS-near data structures in a way that by-passes \omdoc.

Even though IMPS is not in active use anymore, our export is designed to be maintainable so that future extensions can be developed.
This would focus in particular on the export of mid-level proofs and on improvements to the representation of the idiosyncratic parts of the type system such as undefinedness and subtyping.



\section{The Isabelle Library and the AFP}\label{sec:isabelle}
 Isabelle was developed as a logical framework \cite{isabelle2} focusing on automated theorem proving.
Nowadays its most visible use is in the Isabelle/HOL instance, which is used e.g. in the L4 verification \cite{l4verified}.
Isabelle is extremely user-friendly with an out-of-the-box installer and powerful graphical user interface.
It is used widely enough to counter-indicate an integrated library, but formalizations are collected by submission to the Archive of Formal Proof.

Our Isabelle export was carried out together with Makarius Wenzel.
The details are to be published in \cite{KRW:isabelle:19}.

\subsection{Language}

Isabelle uses a very simple higher-order logic called \textsf{Pure} as a logical framework.
Therefore, we did not formalize \textsf{Pure} in LF; instead, we defined \textsf{Pure} as a logical framework in \mmt, i.e., as a sibling to LF.

Within \textsf{Pure}, different logics are formalized, most importantly Isabelle/HOL.
Because the types of Isabelle/HOL are a fragment of the types of \textsf{Pure}, no Church or Curry encoding is used, and HOL functions are directly \textsf{Pure} functions.
This eliminates the scalability hazard we discussed for Coq.

The main difficulty in formalizing the Isabelle language is the module system, which features locales (and as a special case type classes) and morphisms between them.
The foundational details are very subtle and not always obvious.
For better or worse, these are elaborated away almost entirely --- the kernel contains only a few primitive features for extending the \textsf{Pure} logic with type classes that are easily dealt with.
Thus, it is possible to inspect the export to reverse-engineer the details of their treatment.

\subsection{System}

Isabelle provides a heavily-used mechanism for users to define their own high-level declarations, whose semantics is defined by elaboration into more primitive one.
Examples include inductive types and functions.
These high-level statements are invisible to the kernel and lost in the export.
\mmt provides a corresponding mechanism for user-defined high-level features, which we can use to preserve high-level features in the export.
However, as elaboration is implemented directly in ML, this process cannot be automated, and a manual effort is needed for each statement kind to (i) define the corresponding feature in \mmt and (ii) extend the export with code for detecting and exporting its instances.

Similarly, the locale structure is lost entirely in the direct export.
We were able to reconstruct and export the diagram of locales and morphisms non-compositionally.
But because this is an additional data structure, it cannot easily be integrated with exports of the uses of the locales.

\subsection{Libraries}

Isabelle uses a coherent co-maintained standard library, which introduces in particular Isabelle/HOL.
Additionally, user-submitted contributions are collected in the Archive of Formal Proofs library and, once accepted, also co-maintained.

Isabelle is the easiest to work with of all the systems we discuss here, featuring easy installation and a high-level export interface via Isabelle/PIDE.
As part of our export \cite{KRW:isabelle:19}, we funded the main developer to make major upgrades to the PIDE infrastructure, which has led to a very smooth export kernel-instrumenting export module that is fully integrated with Isabelle.
Moreover, because PIDE and \mmt are both written in Scala, we could skip the generation of intermediate files: we built an Isabelle plugin that makes the \mmt library available so that Isabelle can directly pass \mmt data structures to \mmt.
This allows producing OMDoc files directly from Isabelle, making it the most maintainable and scalable of our exports.

However, even with the more space-efficient representation of expressions, Isabelle proof objects become very big.
This is because (i) many commonly-used types (in particular inductive and record types) are elaborated, (ii) Isabelle's strong integration of automated provers  makes it easy for users to build large proofs, and (iii) the lack of implicit computation requires elaborating all computations into equational reasoning.
Therefore, we have opted for a dependency-only export of proofs.

\subsection{Outlook and Open Challenges}

The Isabelle export is extremely well-maintainable as most of it was written by a core developer and deeply integrated with the Isabelle system and code base.
The two biggest future challenges will be the avoiding elaboration and reducing the size of proofs terms.

Regarding \textbf{elaboration}, Isabelle is special among proof assistants in that it a very complex processing pipeline from heterogeneous user-level declarations to homogeneous kernel-level declarations.
Not only does any export based on instrumenting the kernel lose a lot of valuable structure, it is also very difficult to modify the system to preserve that structure.
The most realistic approach is to (i) modify the Isabelle data structures for user-level declarations in such a way that they must also govern how they can be exported.
Existing language features would then (ii) have to be adapted to this new interface.
This could then be combined with (iii) mimicking the user-level features in \mmt to obtain a structure-preserving export.
While (i) and (iii) are relatively easy, (ii) requires understanding and adapting a lot of code for the individual features --- something that even Isabelle kernel developers like Makarius Wenzel may not be able to do easily.

Regarding the size of \textbf{proofs}, we expect the problem to be even bigger than for Coq because the computation of inductive functions (which is a primitive kernel feature in Coq) is realized via equality reasoning is Isabelle and therefore explicitly part of the proof term.
Moreover, Isabelle's extremely strong integration of automated provers increases the disconnect between user-level and kernel-level proof size: user-level proofs may simply consist of invoking automation, which then produces highly circuitous proofs.
This can be alleviated somewhat by engineering efforts such as enabling structure sharing in the export.
But to obtain an export of mid-level proofs as discussed in Section~\ref{sec:proofexp}, investments into proof simplification may be needed.
On the positive side, Isabelle's Isar language already allows users to write mid-level proof terms explicitly.
Thus, it is promising (but still difficult) to export Isar proof terms enriched with information from kernel-level proofs.



\section{The Mizar Mathematical Library}\label{sec:mizar}
 Mizar is one of the oldest theorem provers and the only major one to be initiated on the Eastern side of the Iron Curtain.
It is set apart from most other provers in multiple ways, which reinforce each other:
It focuses almost exclusively on mathematical content (indeed, common computer science constructs such as inductive types and anonymous functions are absent) and uses first-order set theory as the base logic with a highly idiosyncratic syntax.
It is supported by a relatively small developer and user community that has experienced comparatively little cross-pollination with other prover communities.
Yet, it features a large centrally curated library, and many of its seemingly unusual features have stood the test of time for formalizing mathematics.
Papers on formalizations are generated automatically from Mizar sources and published in a special journal.

Our Mizar export was carried out together with Josef Urban and Mihnea Iancu.
The details were published in \cite{IKRU:mizar:11,IKR:mizar:11}.

\subsection{Language}

Due to its idiosyncratic syntax, the Mizar language is often confusing.
But logically, it is actually very simple: it is a formalization of Tarski-Grothendieck set theory in untyped first-order logic with toplevel second-order binders (which are needed to formalize schemas).
This can be formalized routinely in LF.
Structures/records are realized as functions, whose domain is a special set of names.
This is also easy to formalize although, without Mizar's special syntax support for it, the representation is hard to read.
Mizar allows for a large but fixed set of definition principles.
These were difficult to identify and compile, but then straightforward to represent as high-level \mmt statements \cite{IKR:mizar:11}.

Additionally, Mizar uses a soft type system, where types are defined as unary predicates over sets.
This type system is undecidable, and Mizar provides automation for discharging type-checking obligations, partially guided by user-tagged typing theorems (called registrations).
Because this type system is internally represented as typing axioms/theorems about the terms in untyped first-order logic, the direct representation in LF is also straightforward.

However, we could do better:
Our Mizar export predates \mmt implementations of LF with predicate subtypes.
In a reimplementation, we would investigate representing a Mizar type $T$ as the predicate subtype of the type of sets containing those sets that satisfy $\enc{T}$ (akin to how we used predicate subtypes in the Coq export).
This would allow preserving the type system in the export.
However, in that case, rechecking the export would require reproving the type-checking conditions discharged by Mizar, a non-trivial task.

\subsection{System}

Mizar is one of the oldest proof assistants around and was originally designed in relative isolation.
Moreover, due to its focus on mathematics, the user community has remained much smaller than those of the other systems that can be used for software verification.
Therefore, it is the least accessible of all the provers we worked with.

A Mizar export can be best written based on the XML files generated by Mizar.
In fact, these were introduced by Josef Urban for that purpose \cite{mizar_fol} and later became used internally.
Therefore, they contain all relevant information.
But the XML schema is very complex, not always documented, and may change across Mizar versions.

\subsection{Library}

Our export \cite{IKRU:mizar:11} uses the XML files.
It was relatively easy to build (if a system expert is at hand to explain the XML) and to export the Mizar Mathematical Library, the single coherent library that is co-maintained with Mizar.
But the export has been near impossible to maintain over multiple years.

Mizar uses a big kernel with substantial built-in automation, in particular for the soft type system.
These proof steps are not exported and therefore occur as omitted steps in the export.
Except for these gaps, the proofs are mostly available in the same high-level format in which they were written by the user.

\subsection{Outlook and Open Challenges}

\textbf{Maintainability} is a big challenge for ours and future Mizar exports.
The concepts of the Mizar language, its code base, and the structure of the generated XML files are hard to understand.
This is partially inherent in the system and partially an artefact of the Mizar community overlapping less with over proof assistant communities and the broader computer science community.

A logistical challenge here is that the Mizar developer community is small, mostly concentrated in Bia{\l}ystok, Poland, and not as well-funded for international travel as other prover communities.
The resource-intensive maintenance of the system and the library limits their ability to invest into exports.
While \textbf{funding} is an issue for all proof assistant exports, it is particularly limiting for Mizar that major investments into exports can only be funded via research grants from other researchers interested in using the export.

A promising avenue is the \textbf{reconstruction of the Mizar system in a logical framework} in such a way that the Mizar library can be migrated to the framework and --- in the long run --- the Mizar retired.
The main challenges here are mimicking the highly human-friendly concrete syntax of Mizar and the automated reasoning capabilities.
Promising first results have been accomplished in \cite{mizar_isabelle} on top of Isabelle.
A modern reimplementation on top of an LF-like framework, e.g., implemented in \mmt, could also be promising: it would allow for supporting Mizar's dependent types more naturally but would hardly be able to match Isabelle's automated reasoning.

A \textbf{proof export} from Mizar is generally difficult because not all proofs are currently stored in the system.
Because the Mizar kernel is large and complex, instrumenting it for full proof export would require a major effort.
However, an export of mid-level proofs in the sense of Section~\ref{sec:proofexp} is already possible although we did not focus on it.
A related problem is that Mizar elaboration expands many abbreviations including for basic symbols like implication.
Both instrumenting these expansions during or reintroducing the abbreviation after an export are very difficult, and only a few qualified system experts exist for such tasks.
A reconstruction inside Isabelle would allow reducing those problems to the existing Isabelle export technology.



\section{The PVS Prelude and the NASA Library}\label{sec:pvs}
 PVS is a proof assistant based on higher-order logic with powerful additional features inspired by mathematics (classical logic, subtypes), the heterogeneous method (theories and morphisms), and proof automation (in particular decision procedures).
Using an unusual base logic and being developed by a relatively small group of people at a Californian research institute, its user community and publication footprint are smaller than and overlap less with those of HOL Light, Isabelle, and Coq.
But this is not indicative of its quality, and it is used in many research projects, most prominently by NASA, which maintains the largest PVS library.

Our PVS export was carried out together with Sam Owre, Natarajan Shankar, and Dennis M\"uller.
The details were published in \cite{KMOR:pvs:17}.

\subsection{Language}

PVS uses higher-order logic for which a Church-encoding works well.
But it adds major additional features that go beyond what can be formalized in any declarative logical framework.
Therefore, we extended LF with appropriate features as described below.

\paragraph{Type System}
PVS uses predicate subtypes.
Thus, the typing relation is undecidable and therefore cannot be represented in a Church-encoding in a framework with decidable typing.
Similarly, it introduces subtyping, which is highly impractical to represent in a Church encoding.
It would require a non-compositional transformation that introduces names for every \emph{occurrence} of a predicate subtype (applying closures if within the scope of binders) and inserts a named injection functions for every application of subtyping.
And because it is undecidable whether tho predicate subtypes are equal, these injections would have to be generated gradually whenever PVS proves a subtype relation.

Instead, we extended LF with predicate subtypes (which we already used for Coq already).
This makes it possible to formalize PVS's predicate subtypes in a straightforward way while maintaining the advantages of using a Church encoding.
We inferred the additional arguments required for the Church representation in the logical framework, which turned out to be just feasible for a dependency-only export of the PVS libraries.
Alternatively, we could have used the weaker Curry-encoding as for Coq.

\paragraph{Module System}
PVS uses an unusual module system.
Theories may carry parameters, and theories may include multiple instances of the same theory with different arguments, e.g., to include lists over booleans and lists over integers.
All these includes use the same included identifiers, and PVS disambiguates for each occurrence of an identifiers which instance it belongs to.
Consequently, internally all included identifiers carry a long list of implicit arguments that instantiate the parameters.

There is a subtle problem here:  As not every identifier depends on all parameters of a theory, it is possible that identifiers included via different instances are actually equal.
As that equality is undecidable, it becomes undecidable which set of identifiers has been included.
Therefore, PVS's parametric includes are very difficult to eliminate in the export.
Therefore, we chose to mimic them by adding a special include-declaration that abstracts over all parameters of a theory.

Finally, PVS uses primitive (co)inductive types and elaborates them into an axiomatization.
That makes it possible to skip them and to export the elaborated statements only.
But we still had to use some extensions of LF to represent the associated recursive functions.

\subsection{System}

PVS includes a very well-built export facility based on kernel-generated XML files written by Sam Owre.
This is intended specifically for exports like our as well as knowledge management applications like search.
PVS was the only system encountered that already had such a system-near export facility specifically designed for that purpose by the developers.

Because PVS uses a big kernel, very little elaboration takes place.
Therefore, the XML files include all user-level information enriched with the information inferred by the kernel.
This includes virtually the entire original source structure, down to source references and redundant brackets.

Because the type system in undecidable, every statement generates some type-checking-conditions, which must be proved separately.
After proving these, PVS inserts them before the respective statement.
Exports can choose to include or skip these additional theorems.

\subsection{Libraries}

PVS uses a small standard library, called the PVS prelude.
Other libraries are distributed, but only the big NASA library is of major use.
PVS releases are regression-tested against this library.

Our export \cite{KMOR:pvs:17} used the generated XML files for both libraries.
As a part of this export collaboration, the XML schema was heavily debugged and well-documented.
Concretely, we represented the PVS XML schema in Scala, and then generated \mmt data structures from it directly.
Our Scala implementation of the PVS XML schema is also a valuable resource as an alternative documentation of the PVS language.
The export worked very smoothly (once the theoretical challenges had been resolved).
A minor problem we encountered were that the source files in the NASA library must be built in a specific order, a process that PVS does not automate well yet.

PVS stores high-level proof objects in separate files.
But the big prover kernel uses a lot of automation that is not recorded in these proof objects, in particular the heavy use of decision procedures.
Thus, proofs have significant gaps.

\subsection{Outlook and Open Challenges}

PVS supports a very robust and well-documented export to XML that includes all user-level declarations.
That makes it relatively easy to build and maintain exports.

The biggest challenge is that PVS uses a few \textbf{unusual language features}.
This includes (co-)inductive types and recursion at a similar level of difficulty as in Coq.
(Coq uses dependent types and PVS uses predicate subtypes; neither subsumes the other, but both cause similar difficulties.)
But it also includes an unusual kind of includes between theories, theory parameters representing theory morphisms, anonymous record types, and operators for function/record updates.
While these features are quite justifiable, their subtleties make representing the language in a logical framework difficult.
We were able to do so by extending the LF with corresponding features so that transporting the library to other systems remains difficult.

Just like for Mizar, a complete \textbf{export of proofs} is not possible as many proof steps are performed automatically by a large kernel and not tracked.
However, if an appropriate standard as discussed in Section~\ref{sec:proofexp} exists, an export of mid-level proof terms can be realized based on XML files containing partial proof objects.



\section{Conclusion}\label{sec:concl}
 We have presented experiences from building exports of major theorem prover libraries spanning some five years.
We have focused on describing the current state and open challenges in these exports with an eye towards supporting the long term community goal of integrating prover libraries and making provers interoperable.

\subsection{Lessons Learned}

The most important lesson to draw from our work is that prover library exports are possible but very difficult and tedious:
Theoretically, they require state-of-the-art representations of logics in logical frameworks and even the design of new framework features.
Here the development of \mmt, which makes implementing these logical framework feasible, was critical.
Indeed the exports motivated and drove the development of new features in \mmt. 

Practically, each export required enormous amounts of work both from us and from a dedicated collaborator on the system side.
Each export runs into numerous rare and/or undocumented subtleties that require extensive communication between these two.
Here careful planning was critical to to avoid running out of time, personnel, or funding half way through.
Because of this high threshold to have working exports at all, many added value services and system interoperability solutions that would otherwise be in reach (albeit still subject to substantial to research efforts) can hardly be attempted at all.
With our six working exports this work can commence now, and indeed we are in contact with multiple groups that are already using them. 

It is also difficult to rigorously represent the syntax and semantics of the logics of practical provers.
Formal descriptions of the exact logics realized in the provers are typically incomplete, outdated, or distributed over multiple publications.
In particular it is virtually impossible to clarify all details without talking to a system expert.
We even routinely resorted to reverse-engineering the internals of the kernels to find out how corner cases are actually treated.
It is even harder to then represent these logics in logical frameworks.
While the latter are excellent for textbook logics such as FOL, HOL, or pure type systems, the representations of all features used in practical provers go beyond their current expressive power.
The use of \mmt to flexibly extend the framework was critical here, but even so the task remains very difficult.
But the investment is well-spent, as the formalization can serve as an exact, consolidated, and up-to-date documentation of the details and semantics of the logic actually used in the system. 

Because of the above, it is important that exports always preserve all details of the prover's logic and refrain from mixing the export with non-compositional transformations to eliminate complex or idiosyncratic features of the logic.
The latter may sometimes be necessary, especially to integrate libraries across provers, but the question of how to it is in itself a difficult research question that must be separated from the engineering question of realizing any export at all.

Finally, while we have not touched on logic translations and library integration in this paper --- that is a task that builds on top of the exports we have discussed --- we want to point out a misconception that we suffered from ourselves 10 years ago and that we have encountered among many colleagues since: successful library integration will not be based on logic translations.
For example, to translate the Coq library to PVS it is neither necessary nor sufficient to give a formal translation from Coq's calculus of inductive constructions to PVS's higher-order logic.
The real difficulty would be to match up a formalization of, say, the algebraic hierarchy in Coq with that of PVS.
We predict this will work best by abstracting to an intermediate logic strong enough to formalize algebra but weak enough to embed into both Coq and PVS.
Initial attempts towards this kind of aligning libraries in different logics are being made by multiple groups now \cite{dedukti_sharing,KKMR:alignments:16,align_kaliszyk}.

\subsection{Future Work}

We remain convinced that the approach described in this article is without alternative as any interoperability between proof assistants will require library exports of some kind.
In the long run, the community must organize and invest efficiently the resources needed for prover developers to build and maintain such exports.
A standardization of the interface format, as suggested by OMDoc/\mmt, will be helpful, but further research is needed to design these formats, especially for high-level declarations and proofs.

Major efforts will be needed to export libraries of more provers.
Because of resource constraints, we had to choose representative prover libraries so far, and some like Nurpl, Lean, and HOL4 are still prominently absent at this point.
Note that logical similarly or even equivalence such as between Coq and Lean or HOL Light and HOL4 is not the most relevant indication for the challenges faced by exports.
While logical issues such as the representation of the type system in the logical framework are indeed similar, the biggest challenges are often in the engineering where equivalent systems often differ drastically.
Indeed, perceived engineering improvements are often the reason behind system forks and re-implementations. 
For example, a Lean export would likely be entirely different from our Coq export whereas an export HOL4 could be reasonably to our HOL Light export.
In addition to what we presented here, we have so far developed additional (sometimes partial) exports for some provers including TPS, Specware, and MetaMath as well as for a few computer algebra systems; while we have not discussed those here, our conclusions apply to and are informed by them as well.

\subsection{Recommendations for Prover and Library Development}

We have a single clear and important recommendation for developers of theorem provers and libraries: \emph{design systems with exports in mind!}

Scalable and maintainable exports require some design decisions that are not obvious in the beginning because they often do not affect the remainder of the system.
Exports typically only become interesting once a large library has been built, and at that point retrofitting export architectures to mature systems usually presents substantial difficulties.
On the positive side, integrating these design recommendations into new systems is relatively easy and has overall positive side effects outside the export facility.
Concretely, we recommend the following.

Provers should \emph{trace user-level content} on its way through the elaboration pipeline into the kernel-level syntax.
The user-level statements should remain accessible even after the corresponding kernel-level statements have been processed, and the latter should point to the former.
Moreover, these cross-references should be fine-granular, ideally at the level of every single subexpression that occurs in a kernel-level statement.
This would in particular ensure the availability of source references everywhere.
Realizing this may however impact performance substantially and therefore must be carefully considered as a part of the overall design problem.

Provers should commit to \emph{maintaining a system-near export} in some standard data format such as XML or JSON and with a well-documented schema.
Rereading this export should be part of regression test suites.
More generally, a kernel instrumenting interface can be provided.
But even then a default instrumentation should be provided that builds a system-near export.
The system-near export should contain all information relevant for third-party processing, ideally enough to reconstruct the original source for each fragment.
If elaboration is traced as described above, it should in particular include the user-level statements.

Provers should \emph{allow users to structure formalizations} with library integration in mind.
Most prover logics are much more powerful than needed for most concrete formalizations.
For example, the vast majority of the Coq library only uses a few universe levels, but Coq allows an arbitrary number.
That is a perfectly reasonable design for provers, but it makes it harder to reuse formalization in other, less or differently expressive logics.
It is generally difficult to retroactively check which logic features were used in a formalization, and even then it is often the case that the strength of the underlying logic has unintentionally leaked into the formalization.
If a particular formalization only requires, for example, monadic second-order logic, users should be able to limit the strength of the logic in such a way that the kernel checks that the formalization stays within the intended fragment.
That would allow users to control how well their formalizations, once exported, can be reused in other systems.
Realizing such a feature in new provers and using it in new libraries is relatively easy, but retrofitting it to existing ones may be very difficult.


\begin{acknowledgements}
Even though the colleagues with whom we worked on the exports (see the co-authors of the cited papers) were not directly involved in this paper, our discussions with them at the time have influenced this paper as well.
We gratefully acknowledge project support by the German Research Council (DFG) under grants KO 2428/13-1 and RA 18723/1-1 and from the European Union under Project OpenDreamKit. 
\end{acknowledgements}

\providecommand\seen{seen } \providecommand\selfedit{}
  \providecommand\webpageat{web page at } \providecommand\homepageat{home page
  at } \providecommand\projectpageat{project page at }
  \providecommand\systempageat{system home page at }
  \providecommand\svnrepoat{Subversion repository at }
  \providecommand\January{January} \providecommand\February{February}
  \providecommand\Feb{February} \providecommand\March{March}
  \providecommand\April{April} \providecommand\May{May}
  \providecommand\June{June} \providecommand\July{July}
  \providecommand\August{August} \providecommand\September{September}
  \providecommand\October{October} \providecommand\November{November}
  \providecommand\December{December} \providecommand\AUSTRALIA{Australia}
  \providecommand\ROMANIA{Romania} \providecommand\MEXICO{Mexico}
  \providecommand\ITALY{Italy} \providecommand\USA{USA}
  \providecommand\IRELAND{Ireland} \providecommand\HUNGARY{Hungary}
  \providecommand\JAPAN{Japan} \providecommand\CANADA{Canada}
  \providecommand\SPAIN{Spain} \providecommand\NETHERLANDS{Netherlands}
  \providecommand\UK{UK} \providecommand\SWEDEN{Sweden}
  \providecommand\GERMANY{Germany} \providecommand\openmath{OpenMath}
  \providecommand\fc{forthcoming} \providecommand\PROC{Proceedings}
  \providecommand\omdoc{OMDoc} \providecommand\activemath{ActiveMath}
  \hyphenation{Wiki-Sym}


\end{document}